\documentclass[referee]{aa}
\usepackage{graphicx}
\begin{document}
   \title{Orbital Phase Spectroscopy of GX 301--2 with RXTE-PCA}

   \titlerunning{Orbital Spectroscopy of GX 301--2}

   \author{U. Mukherjee
          \inst{}
          \and
          B. Paul\inst{}
          }

   \offprints{U. Mukherjee}

   \institute{Department of Astronomy $\&$ Astrophysics, 
              Tata Institute of Fundamental Research,  
              Homi Bhabha Road, Colaba, Mumbai--400 005, 
              India  \\
              \email{uddipan@tifr.res.in, bpaul@tifr.res.in} 
             }

   \date{ }

  \abstract{
We have investigated the orbital phase dependence of the X-ray spectrum
of the High Mass X-ray Binary pulsar GX 301--2. Here we present the results from
a spectral analysis of two sets of observations of GX 301--2 with the
Rossi X-ray Timing Explorer (RXTE). Of particular interest are the variations
of the absorption column density and the iron line flux along with other
parameters of the spectral model with the orbital phase. We found that
the X-ray spectrum can almost always be fitted with a partial covering
absorption model. We have detected enhanced absorption near the periastron.
However, the column density variation with orbital phase is not smooth,
as is expected in a smooth stellar wind model. 
We discuss the results of the column density variation in the light of the two
proposed models for GX 301-2, an equatorial disk emanating from the
companion star Wray 977 and a gas stream following the neutron star.
We also found that the iron K${\mathrm \alpha}$ and K${\mathrm \beta}$ line fluxes have 
peaks near the periastron and are well correlated with the continuum hard X-ray
flux. The line equivalent width shows an interesting pattern with the
column density, reasonably constant for low values of the column density
and increasing rapidly beyond a certain value.} 

   \maketitle


\section{Introduction}

In High Mass X-ray Binaries (HMXBs), the mass accretion onto the compact object
can be from both the stellar wind and an outflowing equatorial disk of the
companion star. Variation in the stellar wind accretion rate can give rise to
a smooth modulation in the X-ray intensity with orbital phase, whereas strong
enhancement in X-ray luminosity can be seen following the passage of the compact
star through a circumstellar disk. If the companion star has a circumstellar
disk, the occasional increase of material in the disk or an enhancement in the radial extent
of the disk causes the strong X-ray outbursts in such systems. In many of the
HMXBs, wind accretion plays a major role and the
compact object is X-ray bright throughout the binary
orbit, whereas in some other sources there is very little accretion from
the stellar wind and transient accretion from the disk is more
important. There are however a few systems like GX 301--2, in which
a smoothly varying accretion component from the stellar wind and a rapidly
varying component either from a disk or a trailing gas stream are important
and present in almost all the binary orbits. A majority of the HMXBs
harbour high magnetic field neutron stars causing the X-rays
from the compact object to pulsate. X-ray irradiation and photoionisation
of the stellar wind is very important in HMXBs.
Measurement of the orbital evolution of the spectrum by observations
in multiple orbital phases allows us to understand the wind structure and
the effect of the wind on the X-ray spectrum both in terms of absorption and
changing mass accretion rate. The presence and role of a circumstellar disk
of the companion star in producing the orbital intensity variations in the
light curve and changes in the absorption column density with the binary
phase can also be investigated with multi phase observations of such
systems.

GX 301--2 (4U 1223-62) is an HMXB pulsar (White et al. 1976)
with an eccentric binary orbit. 
Using a 200-d Ariel 5 observation of the X-ray source the orbital
period was estimated to be $\sim41$ days from the pulse arrival time 
delays (White, Mason $\&$ Sanford 1978).
The orbital parameters were in turn refined by Sato et al. (1986) by
combining data from Ariel 5, SAS 3, and Hakucho satellites and subsequently
by Koh et al. (1997) using BATSE. The orbital period determined from extensive
BATSE data is $\sim$ 41.5 days and the eccentricity is $\sim$ 0.46.
The mass of the companion star should be less than 55 ${\mathrm M_{\odot}}$ in 
order not to overflow the Roche lobe at periastron. The companion star
Wray 977 was classified as of spectral type B2 Ia e (Parkes et al. 1980),
already evolved off the main sequence, and is at least $10{^6}$ years old.
The P-Cygni type emission line profiles in its optical spectrum show the
presence of an extended and expanding atmosphere. Based on a comparison
of its optical spectrum with that of $\zeta$ Sco, one of the brightest stars
in our galaxy, the companion star was classified to be a B1 Ia+ and the
distance was estimated to be 5.3 kpc (Kaper et al. 1995).

Parkes et al. (1980) evaluated the radius of the companion $R_{\mathrm c}$ 
$\sim$ 0.30a and the Roche radii $R_{\mathrm L}$ $\sim$ 0.64a where `a'
is the semi-major axis of the elliptical orbit. Since Wray 977 substantially
under-fills its Roche Lobe, the dominant mechanism powering the X-ray source
for this system is thought to be stellar wind accretion. 
The mass-loss rate for Wray 977 is in the range of (3--10) $\times$
$10^{-6}$ ${\mathrm M_{\odot}}$ ${\mathrm yr^{-1}}$ (Parkes et al. 1980). 
The stellar wind velocity measured at a distance of $\frac {r}{R_c}$ $\sim$ 
3 from the hydrogen Balmer lines is $\sim$ 300 ${\mathrm km}$ ${\mathrm s^{-1}}$ 
(Parkes et al. 1980). This suggests a terminal wind velocity ($v_{\infty}$) 
of $\sim$ 400 ${\mathrm km}$ ${\mathrm s^{-1}}$, very low for such a star. This large mass-loss rate 
from the companion and the unusually low wind velocity result in stagnation of 
lumps of matter, and a large absorption column density is created along the line of sight.
GX 301--2 shows a variable X-ray luminosity in the range (2-400) $\times$
$10^{35}$ ${\mathrm erg}$ ${\mathrm s^{-1}}$, which depends on the amount of the stellar wind
captured. This in turn depends on the density and velocity of the wind and
hence provides a good site for probing the wind structure.

A consequence of the high eccentricity and changing orbital separation
is a periodically variable accretion rate onto the neutron star.
This X-ray binary shows strong periodic X-ray outbursts before most
periastron passages. Simple stellar wind models cannot account for the
X-ray intensity and absorption column density variations with orbital
phase (White \& Swank 1984) nor can tidal mass transfer (Layton et al. 1998).
However, as shown by Stevens (1988), Haberl (1991) and Leahy (1991), the 
dynamical effects of the neutron star on the stellar wind can highly
increase the mass-loss rate of the companion toward the neutron star near
periastron and may result in a gas stream following it. For GX 301--2
Pravdo et al. (1995) noted a periodic near-apastron flare with lower
intensity than the pre-periastron flare and attributed the two flares
due to accretion from an equatorially enhanced stellar wind or a
circumstellar disk around Wray 977 (also Koh et al. 1997). 

In this work we have investigated the spectral variations of the X-ray
emission throughout the binary orbit. The column density of absorbing
material along the line of sight and flux of the iron emission lines can
be valuable for understanding the wind structure in this system. The X-ray
spectral lines from the X-ray irradiated wind have been investigated in
great detail with the Chandra X-ray observatory (Mukherjee \& Paul 2003,
Watanabe et al. 2003). Here we report the spectral analysis of two sets
of archival RXTE observations covering two orbital periods of GX 301--2.
The observations are described in Sect.2. We then present the variation
of the spectral parameters of the assumed model with orbital phase in
Sect.3 followed by a discussion in Sect.4 and conclude in Sect.5. 

\section{Archival RXTE Observations}

We have used two sets of archival data from 
the Rossi X-ray Timing Explorer (RXTE) 
spanning almost the full orbital period of $\sim$ 41 days.
One set was in 1996 from May 10 to June 15 with seventeen observations
while the other set was in 2000 from October 12 to November 19 with
thirty-nine observations. The 1996 observations did not cover the phase 
interval from 0.85 to 0.98 whereas in 2000 the observations had almost
a full phase coverage. The time of periastron passage is taken to be the
zero of orbital phase (Sato et al., 1986). The useful observation duration
in 1996 was $\sim$ 34 ks and that in 2000 was $\sim$ 262 ks.
 
RXTE consists of a large-area proportional counter array (PCA), consisting
of five Xenon proportional counter units (PCUs) and sensitive
in the energy range of 2--60 ${\mathrm keV}$ with an effective area of 6250 ${\mathrm cm^{2}}$
(Jahoda et al. 1996). It also consists of a high energy crystal
scintillation experiment (HEXTE; 15--200 ${\mathrm keV}$; 1600 ${\mathrm cm^{2}}$ area), and a
continuously scanning all-sky monitor (ASM; 2--10 ${\mathrm keV}$; 90 ${\mathrm cm^{2}}$).
A full description of the ASM detector can be found in Bradt et al. (1993). 
Here we have used the RXTE-ASM long term light curve of GX 301--2 and data
from the above mentioned pointed observations with the PCA. The ASM
lightcurve folded at the orbital period is shown in Fig. 1 with
the PCA count rates superposed on it.
On average three PCUs were on for both the observations. 

\section{Data Analysis and Results}

We took the Standard 2 data products of PCA and extracted the source
spectra for each dataset using the tool saextrct v 4.2d for the appropriate
good time intervals. A systematic error of 1.0\% was added to each
spectral bin. Background data files were generated with the tool
runpcabackest using background models appropriate for the source
brightness and the epoch of the RXTE observation, as provided by the
PCA calibration team. The background subtracted source spectra
were analyzed with the spectral analysis package XSPEC v 11.2.0
(Shafer, Haberl $\&$ Arnaud 1989). Model spectra were fitted to the
observed spectrum for each dataset to determine the important spectral 
parameters.

\subsection{Choice of spectral model}

The broad band X-ray continuum spectra of accreting pulsars are often
found to have the shape of a broken power law or a power law with an
exponential cutoff. The break in the spectrum is in the range of 10--20 ${\mathrm keV}$,
and the power-law photon index below the break energy is in the
range of 0--1 (White, Nagase \& Parmar 1995). Some pulsars also show
cyclotron resonance absorption features, sometimes with more than one
harmonics which are usually modeled as multiplicative Gaussian absorption
components. 
In a few pulsars (e.g, Her X-1, Endo et al. 2000), observations
with good energy resolution CCD detectors showed that the absorption has
two components. One component absorbs the entire spectrum while the other
component absorbs it partially, fitted by the so called partial absorption
model. The partial absorption model is often also described as two
different power-law components with the same photon index but different
normalisations, being absorbed by different column densities. 
In a partial absorption model, the second absorption component should have a size
smaller than the size of the emission region.

Several observations of GX 301--2 with the CCD detectors of the ASCA
satellite clearly showed that the continuum is more complex than a
simple absorbed power-law. The ASCA spectra of GX 301--2 in three different
orbital phases were fitted satisfactorily with a model consisting of a
partially covering power-law component, iron K${\mathrm \alpha}$ and K${\mathrm \beta}$ emission
lines, and an iron absorption edge at 7.1 ${\mathrm keV}$ (Saraswat et al. 1996,
Endo et al. 2002). Using the high resolution X-ray grating spectrum of
GX 301--2 obtained with the Chandra observatory we found that in
addition to the above, there is another broad line component at
6.3 ${\mathrm keV}$, which is the Compton backscattered peak of the bright 6.4 ${\mathrm keV}$ line
(Mukherjee \& Paul 2003, Watanabe et al. 2003). However, the relative
intensity of this line is low and it is too close to the 6.4 ${\mathrm keV}$ line
for inclusion as a separate model component with the moderate resolution
RXTE spectrum. Therefore the spectral model that we have chosen  to
fit the RXTE-PCA spectrum is the same as that of Endo et al. (2002) with
the addition of a high energy exponential cutoff. The latter was not
required with ASCA spectra due to its limited energy band width.

The analytical form of the model that we have used for spectral fitting
is:

\begin{eqnarray}
N(E) & = & {e^{-\sigma(E)N_{\mathrm H1}}}(S_{1}+S_{2}e^{-\sigma(E)N_{\mathrm H2}}+G_1+G_2) \nonumber \\
     &   & {E^{-\Gamma}}{I(E)}{A(E)}
\end{eqnarray}

where
\begin{eqnarray}
\nonumber  I(E) & = & 1  \hspace{0.83in} for ~E < E_\mathrm c \\
\nonumber       & = & e^{- \left({E-E_\mathrm c}\over{E_\mathrm f}\right)} \hspace{0.27in}  for ~E > E_\mathrm c
\end{eqnarray}

and
\begin{eqnarray}
\nonumber A(E) & = & 1 \hspace{0.74in} for ~E < E_\mathrm e \\
\nonumber      & = & e^{-{\tau \left({E_\mathrm e\over E} \right)^3}} \hspace{0.24in} for ~E > E_\mathrm e,
\end{eqnarray}

N(E) is the intensity (no. of photons ${\mathrm s^{-1}}$ ${\mathrm keV^{-1}}$),
$\Gamma$ is the photon index, $N_{\mathrm H1}$ and $N_{\mathrm H2}$ 
are the two equivalent hydrogen column densities,
$\sigma$ is the photo-electric cross-section,
$S_{1}$ and $S_{2}$ are the 
respective normalizations of the power law, $G_1$ and $G_2$ are the
two Gaussian emission lines, $E_{\mathrm c}$ is the cut--off energy,
$E_{\mathrm f}$ the e--folding energy,  
$E_{\mathrm e}$ is the edge energy and $\tau$ is the edge depth.

\subsection{Results}

Apart from a few (seven datasets of 2000 and one of 1996),
almost all datasets gave good reduced $\chi^{2}$ between 0.6 and 1.6
for 44 degrees of freedom. 
The fits with poor reduced $\chi^{2}$ showed wavy residuals with dips
around 10, 20 and 30 ${\mathrm keV}$. Two representative spectra, one with a poor
spectral fit and another with a good spectral fit are shown in
Fig. 2. At this stage we are not sure if this particular nature
of the residuals in some of the spectra is a systematic phenomenon
for this source. Systematic deviations at $\sim$ 20 ${\mathrm keV}$ and $\sim$ 40 ${\mathrm keV}$
were also observed in several broad-band Beppo-SAX spectrum of
GX 301--2 (Orlandini et al. 2000).

Ten spectral parameters were varied in the fitting: $N_{\mathrm H1}$, $N_{\mathrm H2}$,
$\Gamma$, the two normalizations, the two iron line intensities,
edge depth, $E_{\mathrm c}$ and $E_{\mathrm f}$. Of these five parameters
were kept frozen, the iron-line energies
and their FWHM along with the edge energy. For the
parameters which were kept fixed we used the values derived from the Chandra
HETG spectrum (Mukherjee \& Paul 2003). The variations of the free parameters 
with orbital phase are shown in Figures 3 to 6. The 1996 measurements
are shown with filled squares and the 2000 data points are shown with
open triangles. The error bars shown
in the figures correspond to the 90$\%$ confidence levels. For a very few 
spectra near periastron in which the spectral fits were not very good,
the error bars were calculated in XSPEC by increasing the upper limit on the
reduced $\chi^{2}$.

The three spectral parameters defining the intrinsic source continuum;
photon index, cut-off energy and e-folding energy are plotted against
the orbital phase and shown in Fig. 3. Though the total X-ray
luminosity is much higher at the pre-periastron phase, there is no
remarkable change in the intrinsic continuum spectrum. The cut-off
energy varies between 17--22 ${\mathrm keV}$ and the e-folding energy is about
20 ${\mathrm keV}$ except at a few points where the spectral fits are poor.
The variation of the photon index with phase shows that $\Gamma$ in general
lies between 1.0--1.5 for both sets of observations while some values
drop below one. These values for $E_{\mathrm c}$, $E_{\mathrm f}$ and photon index
are more or less consistent with the previously measured values in some
binary phases reported by other authors (White et al. 1983; Orlandini et al. 2000).

The changes of the two column densities with orbital phase for both 
observations are plotted in Fig. 4 (second and third panels).
The figure shows a considerable increase in column density near periastron; but also
a substantial scatter in the values in the intermediate phases. We have
also depicted the variation of the covering fraction measured from our
model in Fig. 5.  From Fig. 6 we find that the iron-line intensities
are much larger than average before the periastron passage and show
a small increase near phase $\sim$ 0.1 for 1996. Though the 2000 set shows a steep
rise near periastron, no such rise near phase $\sim$ 0.1 is noticeable.

\section{Discussion}

\subsection{The Spectrum}

The X-ray spectrum of GX 301--2 is characterised by a heavy and variable
photoelectric absorption with a strong iron K${\mathrm \alpha}$ emission line
(White \& Swank 1984). High resolution observations with X-ray gratings
helped the discovery of a Compton backscattered peak at $\sim$ 6.3 ${\mathrm keV}$ in the
spectrum (Mukherjee \& Paul, 2003, Watanabe et al. 2003). From detailed
Monte-Carlo simulations and comparisons with the observed spectra,
Watanabe et al. (2003) constrained the physical properties of the cold
fluorescing medium surrounding the neutron star. In our analysis
of the low energy resolution RXTE-PCA spectrum covering almost the entire
binary orbit, the 2--30 ${\mathrm keV}$ X-ray spectrum is modeled using a partially covering
high energy cut-off power-law component, two emission lines at 6.4 and 7.0
${\mathrm keV}$, and one absorption edge at 7.1 ${\mathrm keV}$ due to neutral iron. 
The Compton backscattered
peak was not included in this model as the RXTE proportional counters with
moderate energy resolution cannot separate it from the fluorescence line.
We found that this model provides good fits for most datasets throughout
the orbit save for a very few at energies above 8 ${\mathrm keV}$. The reason 
why a few datasets do not follow the general trend remains to be explored.
The average values of the free parameters measured here over the full
binary orbit; viz. photon index, e-folding energy and cut-off energy
follow the general trend which were earlier measured only in some phases
of the binary period (White et al. 1983, Orlandini et al. 2000).  

\subsection{Variations in Column Density}

The equivalent hydrogen column density of material that absorbs the primary
X-ray emission is found to be quite high. At the same time, a large variation
of the column density throughout the binary orbit (from 10$^{22}$ to 
10$^{24}$ ${\mathrm atoms}$ ${\mathrm cm^{-2}}$) seems to be one of the characteristics of the
X-ray spectrum of GX 301--2. The large variation of the column densities
$N_{\mathrm H1}$ and $N_{\mathrm H2}$ at all orbital phases indicate clumpiness of the 
stellar wind at different size scales.

In the partial covering model used to fit the X-ray spectrum of many 
accreting pulsars, $N_{\mathrm H2}$ is interpreted as the column density of the 
material local to the X-ray source, while $N_{\mathrm H1}$ accounts for the rest 
of the material (along with galactic absorption).  
The {\bf covering fraction} is defined as  Norm2/(Norm1+Norm2) 
where Norm1 and Norm2 are 
respectively the normalizations of the two power-laws. 
It is seen from Fig. 5 that the covering fraction remains substantially high almost 
throughout the orbit which means that there is dense and 
clumpy material present throughout. Only a small fraction of the primary X-rays 
come out without facing this dense and clumpy material. The presence of 
a dense cloud around the neutron star and close to it is also
supported by the strong Compton recoil component detected with the Chandra
grating spectrum and its successful reproduction by Monte Carlo simulations 
(Watanabe et al. 2003). 

To compare our measured values of column densities, we have obtained a 
model variation of column density with the orbital phase using a spherically 
symmetric wind emanating from the companion star Wray 977. The stellar wind had 
a Castor, Abbott $\&$ Klein (CAK, 1975) velocity profile :

\begin{equation}
   v_{\mathrm wind} = {v_{\infty}}\sqrt{1-\frac {{R_\mathrm c}}{r}}\,
\end{equation}

where $v_{\infty}$ is the terminal velocity for the 
stellar wind of Wray 977 (400 ${\mathrm km}$ ${\mathrm s^{-1}}$),
R$_{\mathrm c}$ is the radius of Wray 977 
and r is the distance from it.

The column densities were evaluated by  numerical integration along the line 
of sight from the neutron star to the observer at infinity as the neutron 
star traversed the elliptical orbit. The observer was on a diferent plane 
than the orbit and the inclination angle was taken to be 69$^\circ$. 
The projected line of sight made an angle 
of 63$^\circ$ with the major axis of the ellipse [Fig. 6 of Pravdo $\&$ Ghosh 2001 (PG from hereon)]. 

The assumed mass-loss rate was 10$^{-5}$  ${\mathrm M_{\odot}}$ ${\mathrm yr^{-1}}$. 
The result was a smooth curve with a peak 
between phases 0.1 and 0.2 with no rise near periastron (First panel in Fig. 4). 
The values of the column densities were of the order of 10$^{22}$ to 10$^{23}$ 
${\mathrm atoms}$ ${\mathrm cm^{-2}}$
The peak between phases 0.1 and 0.2 was expected since the line of sight passes through the 
densest parts of the wind during these phases. On the other hand, our measured column
densities did not show such a smooth variation and moreover the values 
also varied widely from 10$^{22}$ to 10$^{24}$ ${\mathrm atoms}$ ${\mathrm cm^{-2}}$.  
Thus it is clear that the observed variation in column density cannot be explained 
by a spherically symmetric CAK wind only. 

Two models have been put forward to explain the orbital 
modulation of the column density of GX 301--2. Leahy (1991) fitted the observed 
variation in column density versus orbital phase (from TENMA data) with a spherically 
symmetric stellar wind and a linear gas stream. The peak near periastron was fitted when 
the gas stream was introduced (Fig.3, Leahy 1991). The model with only a CAK wind 
component was found to be unacceptable as it could not fit the rise in the 
column densities at pre--periastron.
A gas stream can be due to the dynamical effect of the neutron star on the companion
wind and the effect would be strongest at periastron when the neutron star is closest to the 
companion. Leahy (2002) also explained the flux peak prior to 
periastron and a broad peak near apastron in the RXTE--ASM lightcurve with the same model.   
However, we note that this is a very simplistic model
since it does not include the viscous time scale of gas flow through the accretion disk 
which is likely to cause a few days delay between the
interaction of the neutron star with the gas stream and the resultant X-ray enhancement.
PG on the other hand, have proposed the
existence of an equatorially enhanced circumstellar disk about Wray 977 which can also
adequately explain the orbital flux modulation of the RXTE--ASM lightcurve. 
The proposed model describes two 
interactions of the neutron star with the disc which gives rise to 
the two peaks in the column density, one at pre-periastron and the other between phases 0.1 
and 0.2, with a smooth component elsewhere. 
However, the orbital dependence of the absorption column density expected 
from such a model (PG, Fig. 7(b)) or for that matter from the gas stream 
model of Leahy (1991) is very different from the variation of the column densities 
at different orbital phases with the RXTE-PCA data as measured by us 
(Fig. 4, second and third panel). 
From the observed variation of the column density 
as measured by us it appears that there are probably strong inhomogeneties in the wind 
that are causing large fluctuations in the column densities. This can be supported by 
the fact that almost throughout the binary orbit the covering fraction remains high (Fig. 5) 
which indicates the presence of clumpy inhomogeneous material. 
  
A possible reason for the increase in column densities  
near periastron may be that 
very strong X-ray emission at the pre-periastron phase ionises most of the X-ray
wind near the neutron star. This in turn can reduce the rate of acceleration of the
wind by the UV emission causing the wind velocity profile to differ from
that of an isolated star. An observable effect of this is additional
clumping of material at the pre--periastron position and an increase in the
measured column densities.

The RXTE-ASM lightcurve of another very similar high mass X-ray binary pulsar
4U 1907+09 was also modeled using a spherical wind model (Roberts et al. 2001).
A phase-locked secondary flare in the lightcurve was reported along with a 
pre-periastron flare. Though the simple wind model fitted the time-averaged 
lightcurve, it failed to account for the secondary flare and also for the
variation of the absorption column density with orbital phase. A spherical
wind plus gas stream in turn fitted the variation in column density.

\subsection{Variation of Iron-line Flux} 

The two iron lines included in our model show large increases in flux
near periastron and a possible small increase near phase 0.1 (at least for 1996).
The peak near periastron (phase $\sim$ 0.9) is not very evident in the
1996 data due to the lack of enough observations, though an increasing trend
can possibly be inferred.
The average magnitude of the 7.0 ${\mathrm keV}$
iron-line flux is about an order of magnitude less than that of the 6.4 ${\mathrm keV}$
iron-line flux. The fluxes for both lines in the intermediate phases
show a more or less steady value. We also found that the
iron-line equivalent width has a clear correlation with $N_{\mathrm H2}$ with 
considerable scatter, as shown in Fig. 7. 
The correlation indicates that
the hard X-rays are reprocessed by the material local to the hard X-ray
source and the neutron star interacts with an enhanced mass near periastron, 
but the presence of enhanced mass near phase 0.1 as proposed by PG is not 
very clearly seen in the line fluxes. 
Previously Makino et al. (1985) and Endo et al. (2002) have showed
that such a correlation exists in GX 301--2 from observations made in certain orbital
phases. Endo et al. (2002) discusses that a simple relation of the
equivalent width exists with the line of sight column density 
as long as the fluorescing plasma is optically thin and fully 
surrounds the pulsar. For that case the equivalent
width increases monotonically with the column density. The scatter seen
in our correlation may be due to the clumpiness of the reprocessing material.   
However, for small values of $N_{\mathrm H2}$ [(2--30)$\times$ 10$^{22}$ ${\mathrm atoms}$ ${\mathrm cm^{-2}}$],
the equivalent width appears to be nearly constant. This may arise if at low
values of $N_{\mathrm H2}$ the clumpy absorber does not completely surround 
the neutron star. Another possibility is that a fraction of the iron line
may be produced by the absorbing material associated with $N_{\mathrm H1}$, 
which becomes the dominant component for small values of $N_{\mathrm H2}$.

\section{Conclusions}

\begin{enumerate}

\item 
The partial covering power-law model with two emission lines and an absorption
edge describes the X-ray spectrum of GX 301--2 well throughout the entire
binary orbit.

\item
The column densities measured are very high with a large variation throughout
the binary orbit indicating a clumpy nature of the stellar wind.

\item
In this work we have measured the variation of column density with orbital
phase and compared it with the variation expected in the disk wind model of
PG  and also in the gas stream model of Leahy (1991).

\item
The correlation of the iron-line equivalent width with the column density
($N_{\mathrm H2}$) suggests that most of the iron line is produced by the local
clumpy matter surrounding the neutron star.

\end{enumerate}

\section{Acknowledgments}
Our work has made use of the data obtained from the High Energy Astrophysics 
Archive Research Center (HEASARC) provided by NASA/GSFC. We express immense 
gratitude towards the anonymous referee for greatly improving our work. 
It is a pleasure for UM to thank K.Garai for his help in wind modeling. 
He also acknowledges the Kanwal Rekhi Scholarship of T.I.F.R. Endowment 
Fund for partial financial support.

\begin{figure}
   \centering
   \includegraphics[angle=-90,width=10cm]{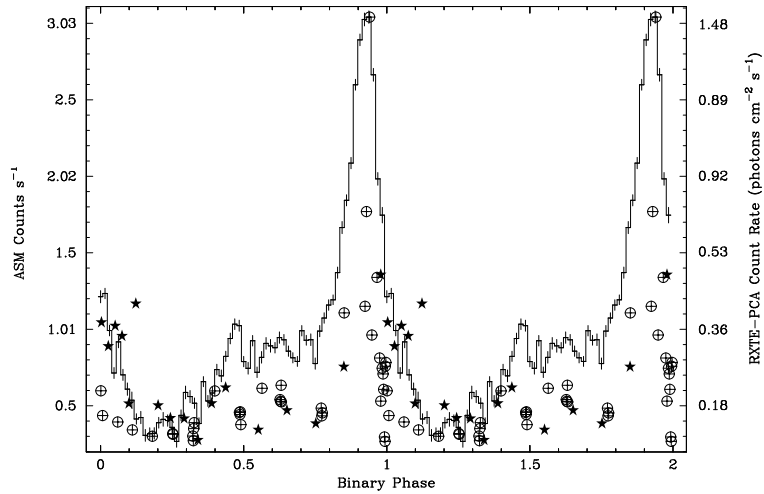}
   \caption{RXTE--PCA count rate superposed on the RXTE-ASM 
period-folded lightcurve as a histogram. The 1996 observations 
are shown with asterisks and the 2000 observations are shown 
with marked circles. The binary phase 1.0 corresponds to 
the periastron passage. 
              }
\end{figure}

\begin{figure}
   \centering
   \includegraphics[angle=0,width=8cm]{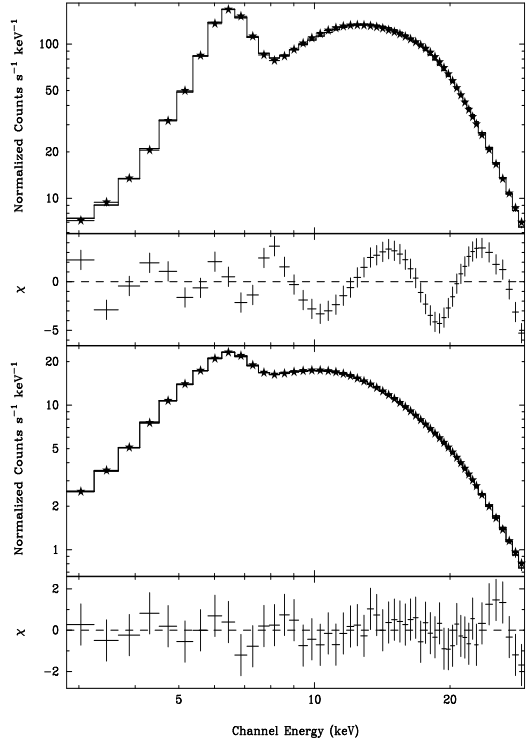}
   \caption{The top panel shows the worst fit spectra 
(ObsID : 50066-02-01-000) with the associated residuals in the 
second panel. The reduced $\chi^{2}$ came out to be $\sim$7. 
The third panel from the top shows one of the many well fitted spectra 
(ObsID : 50066-01-01-00) along with the residuals in the lowest panel.
              }
            \end{figure}

\begin{figure}
   \centering
   \includegraphics[angle=-90,width=7cm]{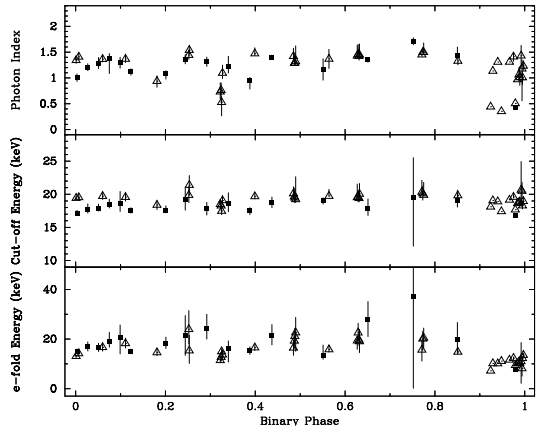}
   \caption{Variation of the photon index, cut-off energy and  e-folding energy 
with the orbital phase in the upper, middle
and lower panels respectively. 
              }
\end{figure}
\begin{figure}
   \centering
   \includegraphics[angle=0,width=16cm]{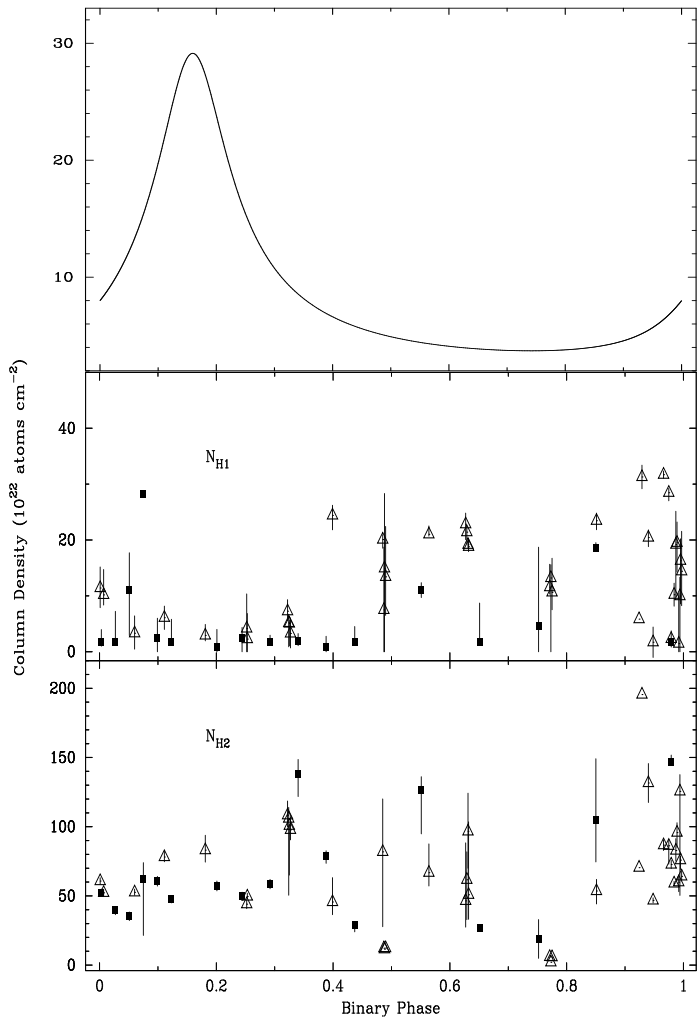}
      \caption{Variation of Column Density with the orbital phase. 
The uppermost panel depicts the model variation, the middle panel 
showing the observed variation of N$_{\mathrm H1}$ and the lower panel
shows the observed variation of N$_{\mathrm H2}$.
              }
\end{figure}
\begin{figure}
   \centering
   \includegraphics[angle=-90,width=10cm]{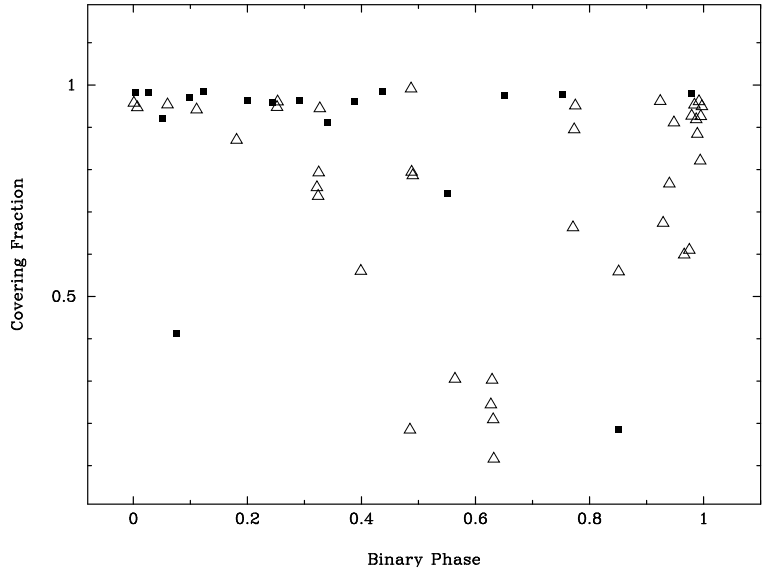}
      \caption{The variation of the covering fraction of the model with the orbital phase. 
              }
\end{figure}
\begin{figure}
   \centering
   \includegraphics[angle=-90,width=12cm]{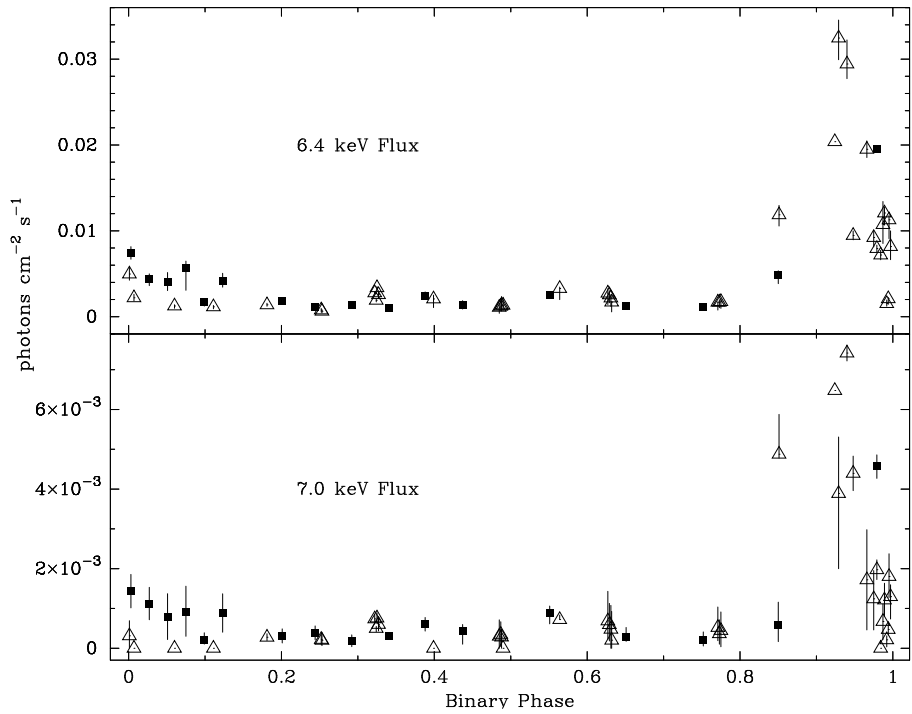}
      \caption{Variation of the iron-line flux of the model with the orbital phase. 
Here the upper panel depicts the flux corresponding to the 6.4 ${\mathrm keV}$ iron-line and the 
lower panel shows the 7.0 ${\mathrm keV}$ Iron-Line Flux respectively.
              }
\end{figure}
\begin{figure}
   \centering
   \includegraphics[angle=-90,width=10cm]{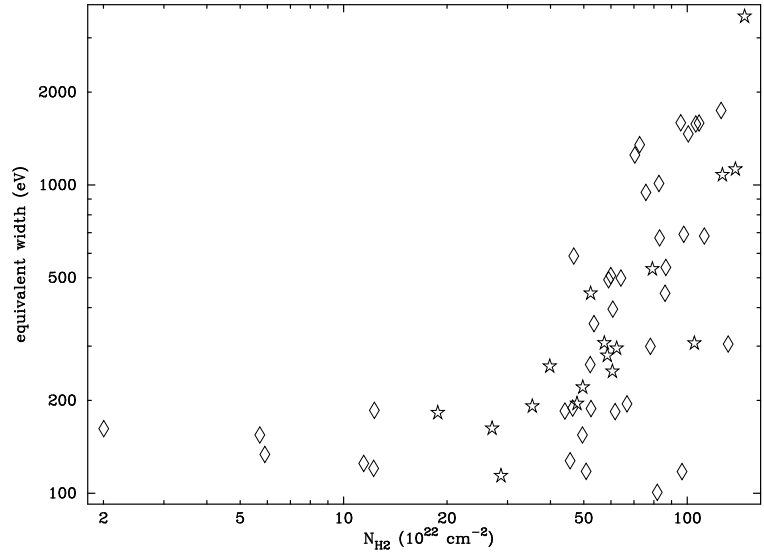}
      \caption{Correlation of equivalent width of the 6.4 ${\mathrm keV}$ iron-line with $N_{H2}$. 
Both the axes are in logarithmic scale. The stars denote the 1996 dataset and the 
diamonds represent the 2000 dataset.  
              }
\end{figure}

\end{document}